\documentclass[acmtog, nonacm=true]{acmart}

\usepackage{subfigure}
\usepackage{amsmath}

\AtBeginDocument{%
  \providecommand\BibTeX{{%
    \normalfont B\kern-0.5em{\scshape i\kern-0.25em b}\kern-0.8em\TeX}}}

\setcopyright{acmcopyright}
\copyrightyear{2009}
\acmConference[]{Department of Computer Science}{April 1, 2009}{University of British Columbia}
\acmYear{2009}
\acmBooktitle{}

\begin{document}

\title{Correction of Chromatic Aberration from a Single Image Using Keypoints}

\author{Benjamin T. Cecchetto}
\email{benjamin.t.cecchetto@gmail.com}
%\email{webmaster@marysville-ohio.com}
\affiliation{%
  \institution{Department of Computer Science, The University of British Columbia}
}

%\renewcommand{\shortauthors}{Cecchett}

\begin{abstract}
In this paper, we propose a method to correct for chromatic aberration in a single photograph. 
Our method replicates what a user would do in a photo editing program to account for this defect. 
We find matching keypoints in each colour channel then align them as a user would.
\end{abstract}



\maketitle

\section{Introduction}
Chromatic aberration (also known as colour fringing) is a phenomena where different wavelengths of light refract through different parts of a lens system. 
Thus, the colour channels may not align as they reach the sensor/film. 
This is most notable in cheaper lenses, it is also noticeable at higher resolutions. 
We desire an image free of chromatic aberrations is simple, so all the planes are in focus. 
For a simple lens system this amounts to misaligned colour channels (red, green and blue). 
The misalignment is due to a uniform scaling and translation. 
There are many types of other chromatic aberrations. 
Complex lens assemblies introduce new distortions.
We propose a method to deal with the simple, more common scenario. 
We later discuss possible ways to deal with more complex ones.

\begin{figure*}[h]
  \centering
  \subfigure[\label{subfig:buildraw}]{\includegraphics[width=0.5\linewidth]{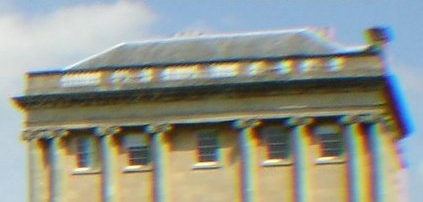}}
  ~
  \subfigure[\label{subfig:buildfixed}]{\includegraphics[width=0.5\linewidth]{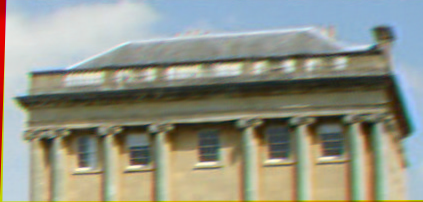}}
  \vspace{-0.5cm}
  \caption{\subref{subfig:buildraw} A photo with chromatic aberration. 
  \subref{subfig:buildfixed} A correction for it using our algorithm.~\cite{wikiaberr} \label{fig:building}}
\end{figure*}

\begin{figure*}[h]
  \centering
  \subfigure[\label{subfig:Lraw}]{\includegraphics[width=0.5\linewidth]{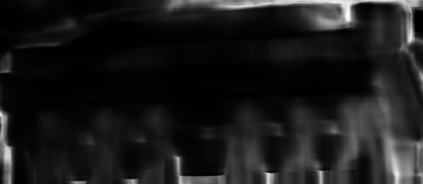}}
  ~
  \subfigure[\label{subfig:Lfixed}]{\includegraphics[width=0.5\linewidth]{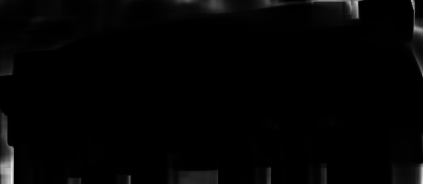}}
  \vspace{-0.5cm}
  \caption{$L$ values over the images in Figure~\ref{fig:building}. \label{fig:Limages}}
\end{figure*}

\begin{figure}[h]
  \centering
  \includegraphics[width=0.5\linewidth]{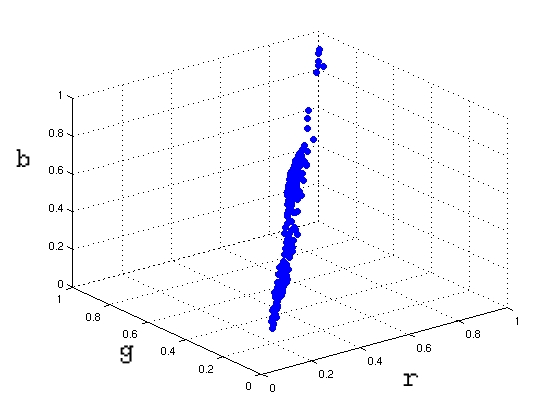}
  ~
  \includegraphics[width=0.5\linewidth]{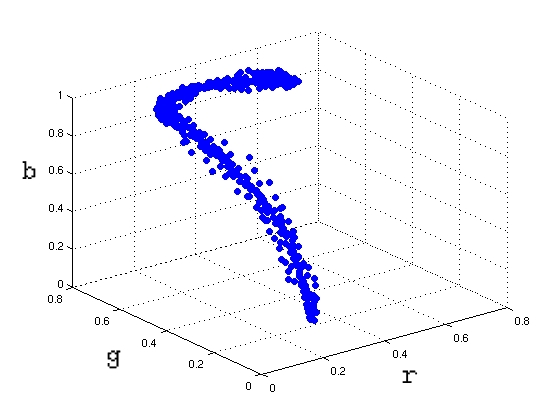}
  
  \includegraphics[width=0.5\linewidth]{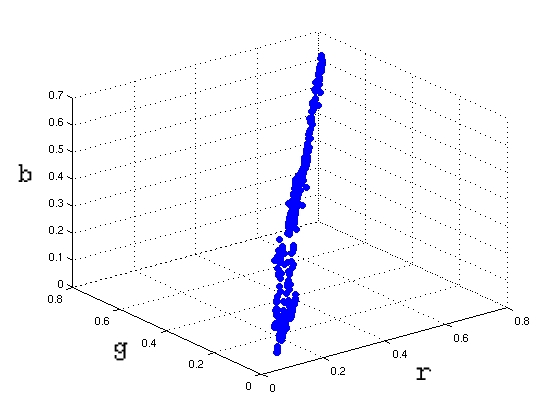}
  ~
  \includegraphics[width=0.5\linewidth]{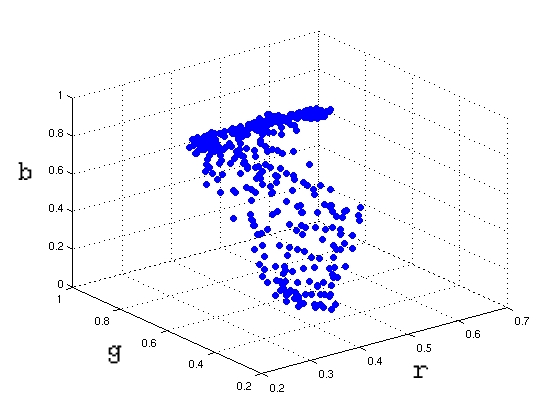}
  
  \includegraphics[width=0.5\linewidth]{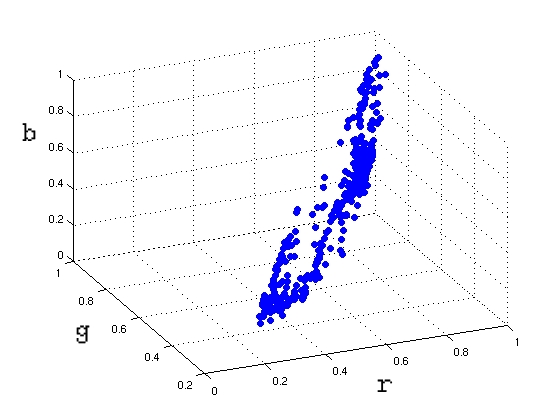}
  ~
  \includegraphics[width=0.5\linewidth]{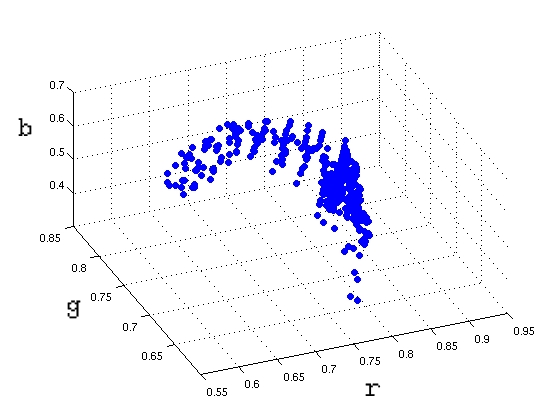}
  
  \caption{Colour points $(r,g,b$) in neighbourhood clusters around a
pixel. 
Left: From a control image with no aberrations chosen randomly.
Right: From a photo with aberrations, chosen at regions of obvious colour fringing.
 \label{fig:colorCurves}}
\end{figure}

\begin{figure}[h]
  \centering
  \subfigure[\label{subfig:cropraw}]{\includegraphics[width=0.5\linewidth]{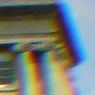}}
  ~
 \subfigure[\label{subfig:cropfixed}]{ \includegraphics[width=0.5\linewidth]{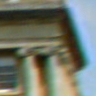}}
 \vspace{-0.5cm}
  \caption{Zoomed in area from Figure~\ref{fig:building}. \subref{subfig:cropraw} Original photo. \subref{subfig:cropfixed} Corrected. \label{fig:crop}}
\end{figure}

\section{Previous Work}
Many domains have sought to correct chromatic aberration. 
A major goal is to improve accuracy of measurement. 
This could be either in a lab or more constrained capture scenarios such as of space. 
One such method is in active optics~\cite{willson1991active}. 
For active optics, instead of  taking one exposure for the colour channels simultaneously, the user takes one for each exposure at different times. 
Each time, the focus changes to match the channel's optimal plane. 
This accounts for the aberration in the lens so the three channels will align in the final result. 
This requires a change from traditional hardware as well as three exposures at three points in time. This limits capture for dynamic scenes, as the object may be moving in that time span.  

Another proposed method is to model the edge displacements with cubic splines to compute a nonlinear warp of the image~\cite{boult1992correcting}. This method cannot handle blurring/defocus in the image plane along with saturation effects. It breaks down in regions that are either underexposed or overexposed.  A different technique exists for correcting lateral chromatic aberration where one calculates the aberrations and uses warping to compensate~\cite{mallon2007calibration}. This only works for calibrated cameras/images which may not be workable for a more simple user. 

Chromatic aberrations can also determine depth information without prior calibration~\cite{garcia2000chromatic}.  Recently, an analysis of aberrations show they are not a simple shift of colour planes. We can get a corrected result by computing a non-linear warp~\cite{kang2007automatic}. This algorithm handles many different categorized cases of aberrations. Its only limitations are artifacts caused by saturation, and slow, non-linear, performance.  

Since we want to deal with the simple case, we may want to remove perspective distortions. A few algorithms have solved this problem in the past~\cite{swaninathan2003perspective}.  If we undistort a lens distortion, we should end up with only needing to solve for the translation/scaling of the colour channels to align them. 

A recently proposed metric, $L$, can determine a depth matte from a colour-filtered aperture~\cite{bando2008extracting}.  This metric measures how collinear and correlated points are in 3D (colour space). It is then used to find a disparity from their custom colour-filtered aperture.  

\section{Overview}

Instead of concentrating on the optical derivation, we instead consider an image-based one. An artist can align colour channels of an image to correct the aberration. They align the channels by moving the corners of the images to transform each channel's image. We attempt to automate this artist driven result. 

The main idea is to align two of the colour channels to the third one. The green channel has the least amount of aberration as it is in the middle of the colour spectrum. In some of the results, we match to the red channel for simplicity. Choosing a different channel to match to results in a scaled version of the corrected image. 

Our solution is to find keypoints in all channels, and find where the keypoints match to the fixed channel. At the same time, we minimize the aberration metric $L$. This is like finding disparities in stereo, but in 3 channels instead of two images. Once we have these matchings, we can prune the results based on how well the matchings are. Then we need to find a transformation from the non-fixed colour planes to the fixed one. We restrict the transformation to a scale and translation. We solve the issues that arise from saturation by ignoring saturated regions and their neighbours. Since the algorithm is linear, it is quite efficient.

\subsection{Computing the Alignment Metric}
Before we start describing how to find the keypoints, and
disparities, it is best to define how aligned the colour channels are. It is not suitable to do cross-correlation, as that only
tells us how aligned two colour channels are. We want to see
how aligned three are. As described above, there exists such
a metric $L$~\cite{bando2008extracting}. For a given neighbourhood
around a pixel $(x, y)$, with eigenvalues of the covariance matrix
trix  $\lambda_i$ and covariance matrix diagonal elements $\sigma_r^2$,$\sigma_g^2$, $\sigma_b^2$,

$$L(x,y) = \frac{\lambda_0 \lambda_1 \lambda_2}{\sigma_r^2 \sigma_g^2 \sigma_b^2}$$
 
This value is essentially how collinear the colour points in
this neighbourhood are in RGB-space. 
These are visualized in Figure~\ref{fig:colorCurves}.
The lower this value
is, the more collinear the points are and the higher the less
collinear. As mentioned in the appendix of the paper, this
can be considered to be related to cross-correlation, and is
thus exactly what we want to use. How to choose this neighbourhood size is a different story. The smaller it is, the less
statistics we have about that neighbourhood and thus may
have a worse matching. The larger it is, the better chance we
have of a matching, however the longer it takes. This value
is bounded between 0 and 1, as mentioned in the paper. If
we show this for every pixel as in Figure~\ref{fig:Limages}, we can see that
images without chromatic aberration show very little misalignment over the whole image.
Another justification of using this metric is that according
to the colour lines model~\cite{omer2004color}, colour
points in RGB-space of the whole image will lie along different colour lines. If we look at smaller neighbourhoods, then
we can assume the points will also lie along either a line, or
intersection of lines. This measures the collinearity of those
neighbourhoods. If we search for an ideal alignment, then
we want to maximize the collinearity, thus minimizing $L$.

\subsection{Finding Keypoints}
The first task is to find regions where the keypoints would be
useful. We want to find regions where the alignment measure
$L$ is very high, but at the same time we want to be certain
there is a good possibility for a good alignment. Choosing
regions in the image with high $L$ values is costly, as we have
to compute $L$ for every pixel and examine that image. Also,
it is not guaranteed to give us a good pixel, such as smooth
regions which may have ambiguous results.
A good choice would be to find high gradient regions and
use points from those. Specifically where we know there is
an edge nearby. This gives us a better match since we know
the other two channels should have a high gradient in that
region too. We randomly sample from the norm of the gradient image with gradient sufficiently high within a threshold.
In addition, if we want to pay the cost of the $L$ image, we
can compute it only in regions where the norm of the gradient is sufficiently high. If we multiply $L$ by the norm of the
gradient image, and threshold it, we can sample points that
are highly unaligned with enough information around them
to be aligned.

\subsection{Finding Disparities}
If we want to align two channels to the remaining one (say
align green and blue to the red channel), we need to shift over
all combinations of possible windows, with different scales.
The idea is we want the neighbourhood of the green and blue
channels to be correlated with the red, and those channels
we know may be elsewhere with a different scale. Thus,
we want to minimize the misalignment $L(x, y)$ subject to a
shifted and scaled window in both green and blue channels
so we can write

$$ L(x,y; d_x^G, d_y^G, \sigma^G, d_x^B, d_y^B, \sigma^B)$$

With disparities $d_x$, $d_y$ and relative scales $\sigma$. Where we iterate over all acceptable disparities (as in stereo) and all acceptable scales.
We know however, that the disparities and
cale difference shouldn’t be too large (unless one has a truely horrible lens) so we can limit the search to local neighbourhoods in that sixtuple. In fact, if the scale difference is
decently small (which it usually is in the case of aberrations)
we can simply look for disparities to find the scale aspect of
the transform. Thus we can write $L$ as
$$ L(x,y; d_x^G, d_y^G, d_x^B, d_y^B) $$

To be able to handle different scales, one could simply
do an image pyramid based approach as in other computer
vision papers. This is unnecessary, as most aberrations are
not that distant in the scale domain.
The reason why this works is if we have a perfect edge (all
one colour on one side, then all the other colour on the other
side of the edge) in an image then the colours in the neighbourhood will cluster into two distinct regions. Since we are
dealing with a natural image and edges are not perfect, these
clusters will connect in a line as there will be a gradient from
one colour to the other. If the image is misaligned, then this
region will become more spread out. Now if we consider
a multicoloured region, we get a more complicated shape.
However, if we find disparities that minimizes this cluster’s
collinearity, we should get a better aligned image since we’re
minimizing the spread of the whole shape.

\subsection{Pruning Keypoints}
Although we’ve found keypoints and disparities for those
keypoints in the other two channels, they might not give us
good information. For example, if the best matching $L$ value
for the point neighbourhood was high, we shouldn’t want
to use that point as it is not a very well aligned point. We
only want to use points that have gone from high $L$ value
(unaligned) to low $L$ value (aligned). Since we only chose
points that are unaligned, we just need to prune away the
points that remain unaligned. Thus we only choose points
with a low enough new $L$ value.
Alternatively, since we know we want to do a scale and
translation for each channel (3 degrees of freedom for each)
we only need 2 keypoints (each being 2 dimensional). Thus
we could choose the 2 keypoints with the lowest $L$ value.
Other methods could include weighting the keypoints based
on the $L$ parameter and haven’t been fully explored. In practice, thresholding by the right amount is sufficient for good
results.

\subsection{Computing Image Transforms}

Now that we have a set of good points, we can solve for
the transformation pretty easily. Let us consider the red and
green channels for now. We have $(p_x^R, p_y^R)$ chosen in the
fixed red channel and a point translated by disparity in the
green channel $(p_x^G, p_y^G)$. We have the equation:

\begin{equation*}
\begin{pmatrix}
\sigma^G & 0 & t_x^G \\
0 & \sigma^G & t_y^G \\
0 & 0 & 1
\end{pmatrix}
\begin{pmatrix}
p_x^R \\
p_y^R \\
1 
\end{pmatrix}
=
\begin{pmatrix}
p_x^G \\
p_y^G \\
1 
\end{pmatrix}
\end{equation*}
Thus we can rearrange it for this point pair as:
\begin{equation*}
\begin{pmatrix}
p_x^R & 0 & 1 \\
p_y^R & 1 & 0 
\end{pmatrix}
\begin{pmatrix}
\sigma^G \\
t_x^G \\
t_y^G 
\end{pmatrix}
=
\begin{pmatrix}
p_x^G \\
p_y^G 
\end{pmatrix}
\end{equation*}

If we have a second point we can solve this system for
$\sigma^G$,$t_x^G$, $t_y^G$, the scale, and translation respectively. In practice, we’d want more than just two points because the points
might have only a good local solution and not global. Thus
we’d want points from different regions of the image.

\section{Results}

\begin{figure*}[p]
  \centering
  \subfigure[\label{subfig:synthraw}]{\includegraphics[width=0.33\linewidth]{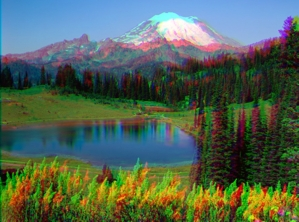}}
  ~
  \subfigure[\label{subfig:synthfixed}]{\includegraphics[width=0.33\linewidth]{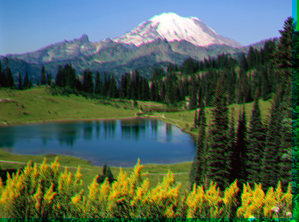}}
  ~
  \subfigure[\label{subfig:synthgt}]{\includegraphics[width=0.33\linewidth]{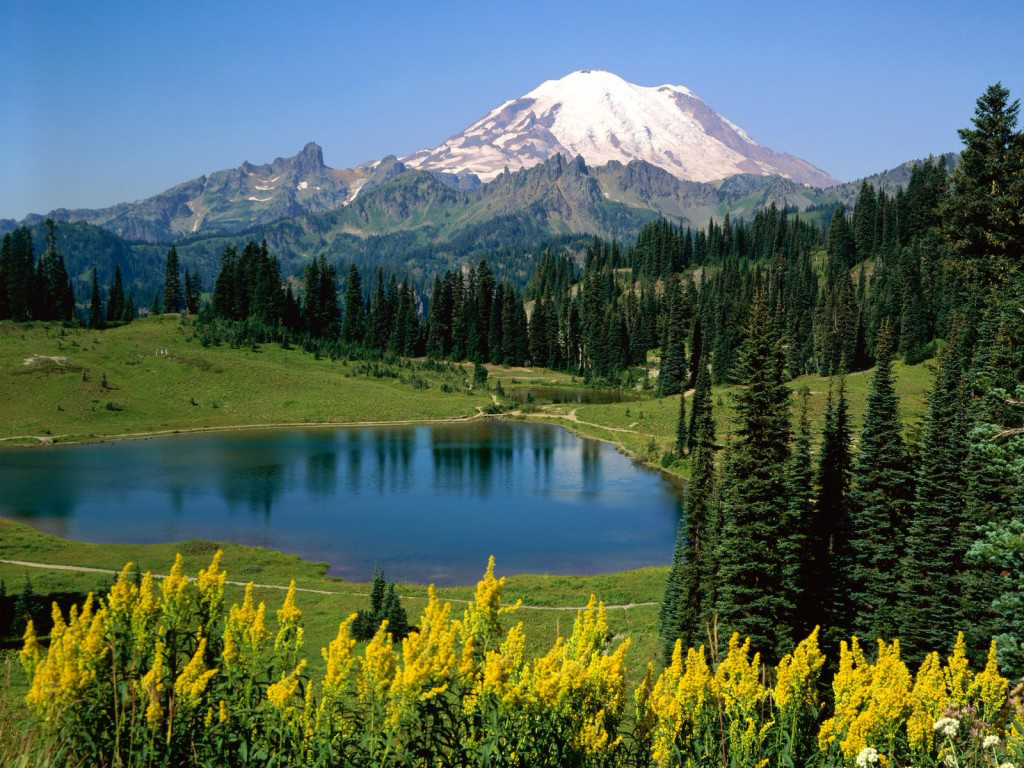}}
  \vspace{-0.5cm}
  \caption{
    \subref{subfig:synthraw} A synthetically aberrated mountain image via translation in different colour channels.  
    \subref{subfig:synthfixed} A corrected version using our algorithm.
    \subref{subfig:synthgt} The ground truth image.\label{fig:transl}}
\end{figure*}

\begin{figure*}[p]
  \centering
  \subfigure[\label{subfig:translraw}]{\includegraphics[width=0.33\linewidth]{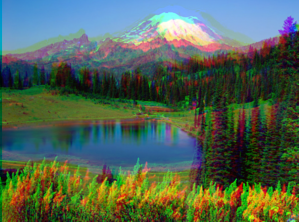}}
  ~
  \subfigure[\label{subfig:translfixed}]{\includegraphics[width=0.33\linewidth]{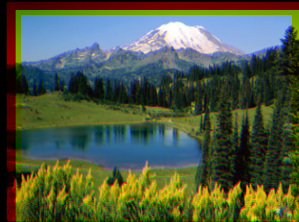}}
  ~
 \subfigure[\label{subfig:translkeypoints}]{ \includegraphics[width=0.33\linewidth]{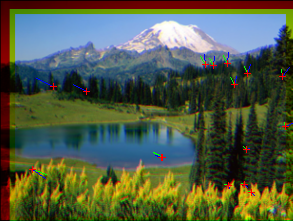}}%
  \vspace{-0.5cm}
  \caption{
    \subref{subfig:translraw} An image with exaggerated translation and scale shift in two colour channels. 
    \subref{subfig:translfixed} A corrected version using our algorithm. 
    \subref{subfig:translkeypoints} The same corrected version showing disparities and keypoints.
    \label{fig:keypoints}}
\end{figure*}

\begin{figure*}[p]
  \centering
  \subfigure[\label{subfig:tripodraw}]{\includegraphics[height=0.18\linewidth]{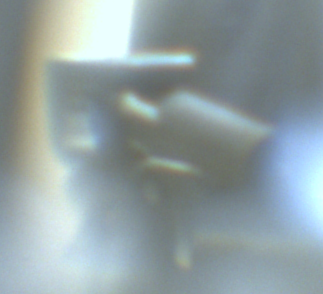}}
  ~
  \subfigure[\label{subfig:boxraw}]{\includegraphics[height=0.18\linewidth]{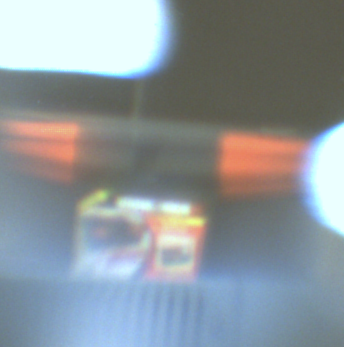}}
  ~
  \subfigure[\label{subfig:limitationraw}]{\includegraphics[height=0.18\linewidth]{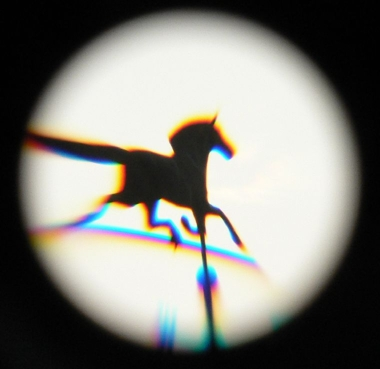}}
  \\
  \subfigure[\label{subfig:tripodfixed}]{\includegraphics[height=0.18\linewidth]{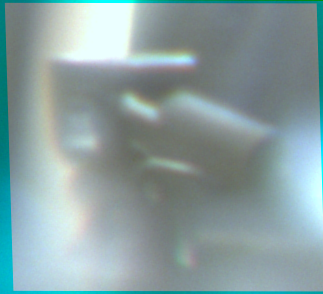}}
  ~
  \subfigure[\label{subfig:boxfixed}]{\includegraphics[height=0.18\linewidth]{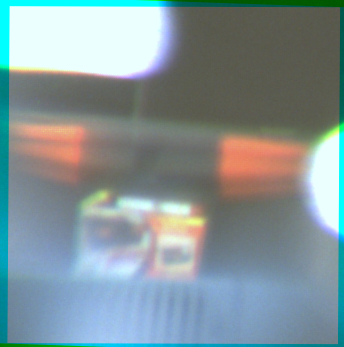}}
  ~
  \subfigure[\label{subfig:limitationfixed}]{\includegraphics[height=0.18\linewidth]{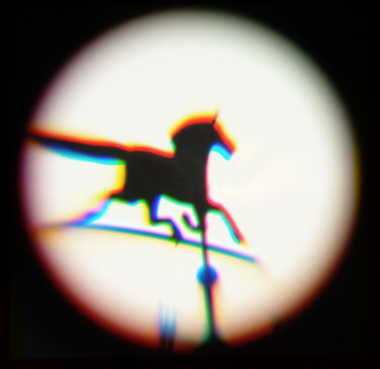}}
  \vspace{-0.5cm}
  \caption{A set of extreme results. \subref{subfig:tripodraw}~An image taken with a lens with large chromatic aberrations. \subref{subfig:tripodfixed}~A corrected version after using our algorithm.
Similarly for \subref{subfig:boxraw} and \subref{subfig:boxfixed}. \subref{subfig:limitationraw} An image with distortion~\cite{zenaberr} and our algorithm result in \subref{subfig:limitationfixed}. Note the properly aligned stand but misaligned horse head.
\label{fig:extreme}
}
\end{figure*}

So far this seems to work quite well as seen in Figure~\ref{fig:building} and
more closely in Figure~\ref{fig:crop}. In
Figure~\ref{fig:transl} we have a synthetic image being corrected. As we
can see the correction is quite similar to the original image.
A lot of the high frequency details are a little blurred because
the misaligned image has a lower resolution than the original
image and thus we cannot get those details back.
In Figure~\ref{fig:keypoints} we have a larger scale change between the different channels. Our algorithm solves this pretty well too.
We can see the keypoints in the third image of this figure and
some matchings aren’t always correct such as the one in the
bottom left pointing in the wrong direction.
In Figure~\ref{fig:extreme}, there are images with extreme lens distortion where our algorithm is expected to fail.
Some of these include photos taken with a lens assembly that exaggerates chromatic aberrations. These photos are blurry because it is very hard to focus with this assembly. 
On the top we have the original photos and the bottom we have the corrected versions. 
Notice that the box is actually worse than the original whereas the tripod has a slight improvement. 
These photos are hard to deal with because the aberration is not as
apparent since all of the colour channels are blurry.

\section{Discussion}

If we reduce the neighbourhood size to compute $L$, we get
more false-positives in the correlations. This is true because
more disparities give a low number. The colours will cluster
into a spherical region in a smooth neighbourhood, whereas
we want lines. It will just choose a random disparity in this
case. For the blurry images in Figure~\ref{fig:extreme}, this is a similar phenomena and there isn’t enough information in the image for
this method to work well and reliably. Many disparities in
this image regardless of direction gave $L$ a value below 0.01
where less than that is considered a ’good’ alignment in regions with more detail.

It was mentioned earlier that we may weight the keypoints
and their appropriate disparities based on how much they reduced the alignment measure. Each row of equation 2 would
be multiplied by a function of its associated $L$. Different
linear and squared error weighting based on $L$ have been attempted with little change in results. One could try to normalize the weights somehow instead of just using $L$ directly.

Also, there might be other statistical methods to explore to prune the keypoints such as computing the translation/scaling and getting rid of the outliers using RANSAC.
The computationally expensive part of this algorithm is determining the disparities where we have a 4D loop (without
handling extremely large scale changes). After we have the
disparities it’s much quicker to deal with the keypoint data,
especially since we need so few keypoints.

Another option to consider speedups is perhaps a hierarchical approach. One could try to solve the problem with a
reduced resolution image and gradually work our way to the
full resolution image using the lower resolution information.

What isn’t clear but is worth exploring is a statement made
earlier saying that unwarping a distorted image will yield an
image with a chromatic aberration that can be corrected with
a scale/translation. This is worth exploring, as the undistortions relatively fast as well as this algorithm.

Another approach to find keypoints would be to segment
the image into many cells and pick an appropriate keypoint
from each as in section 3.3. This will allow enough information from different parts of the image to make a global
warp more accurate than just choosing random points in the
acceptable regions.

One other thing that has been explored is doing the same
thing in gradient domain. Some initial results have shown
that it didn’t work as well as the natural image formations as
in our figures. It is worth exploring trying to align the gradients in a different way, perhaps using chamfer alignment on
the edges of the channels.

\section{Conclusion}
We have presented an method that corrects chromatic aberrations in a single image without any use of calibration. Also, since this method is keypoint based and linear, the method is efficient. Our method also handles the case of saturation, since it can ignore those regions by not choosing keypoints near them.

\begin{acks}
Wolfgang Heidrich for the idea for the project. Gordon Wetzstein for his lens.
\end{acks}

\bibliographystyle{ACM-Reference-Format}
\bibliography{bibliography}

\end{document}